\documentclass[11pt]{article}

\usepackage{graphicx,fullpage}
\usepackage{epstopdf, epsfig}
\usepackage{placeins} 
\usepackage[toc]{appendix}
\usepackage{amsmath,amssymb,eufrak,physics}
\usepackage{bigints}
\usepackage[table,dvipsnames]{xcolor}
\usepackage{upgreek}
\usepackage{soul}
\usepackage{bm}
\usepackage{booktabs,diagbox}
\usepackage{colortbl}
\usepackage[square,numbers,compress]{natbib}

\newcolumntype{C}[1]{>{\centering\arraybackslash}m{#1}}

\numberwithin{equation}{section}
\providecommand{\noopsort}[1]{}

\newcommand{\eq}[2]{\begin{gather}#2  \label{#1} \end{gather}}
\newcommand{\nn}{\nonumber\\}%\label{#1}}
\newcommand{\DD}{\mathrm{D}}
\newcommand{\Gr}{\cellcolor{gray!40}}
\newcommand{\Bl}{\cellcolor{blue!25}}
\newcommand{\Ve}{\cellcolor{green!25}}

\begin{document}

%\title
%{Instability of annular (two-phase) flow
%in a vertical plane channel}
\title{\bf Instability of layered two-phase flow\\ in a vertical plane channel}

% NOTE: A full address must be provided: department, university/institution, town/city, zipcode/postcode, country.
%\makeatletter
%\newcommand{\corresaus}[1]{%
%  \gdef\@corresau{\textbf{Corresponding author:} #1}%
%}
%\makeatother
%\corresaus{Vittorio Di Federico, vittorio.difederico@unibo.it}
\author{{\bf Antonio Barletta}$^1$, {\bf Michele Celli}$^1$, {\bf Vittorio Di Federico}$^2$, {\bf Sandro Longo}$^3$\\[5pt]
$^1${\small Department of Industrial Engineering. Alma Mater Studiorum Universit\`a di Bologna.}\\ 
{\small Viale Risorgimento 2 - 40136 Bologna - Italy}\\[2pt]
$^2${\small Department of Civil, Chemical, Environmental, and Materials Engineering.}\\ {\small Alma Mater Studiorum Universit\`a di Bologna.}\\ {\small Viale Risorgimento 2 - 40136 Bologna - Italy}\\[2pt]
$^3${\small Department of Engineering and Architecture. University of Parma.}\\ 
{\small Parco Area delle Scienze 181/A - 43124 Parma - Italy}
}

\date{}

\maketitle

\begin{abstract}
The equilibrium two-phase flows in a vertical plane channel are explored by adopting an approximate Boussinesq-like model. The class of stationary parallel flows examined in this paper features three-layer configurations in which the central core is less dense than the peripheral layers. Such flows are considered purely buoyant, induced by the density difference between the phases. The simplified scheme leads to a formulation where the parameters driving the transition to linear instability of the equilibrium flow are the Archimedes number and the dimensionless positions of the interfaces between the phases. The analysis takes into consideration symmetric and asymmetric flows, including, as special cases, also two-layer configurations with one phase close to the left-hand boundary wall and the other close to the right-hand wall. In particular, the completely antisymmetric flow, where the equilibrium velocity profile is an odd function of the transverse coordinate, shows up as the least stable configuration among those examined in this study.
\end{abstract}

%
%%%%%%%%%%%%%%%%%%%%%%%%%
\section{Introduction}
\label{sec:introduction}
%%%%%%%%%%%%%%%%%%%%%%%%%
%

Buoyancy-driven exchange flow in a vertical conduit is characterised by the concurrent upward motion of a less dense fluid and the downward motion of a denser fluid. When the conduit connects closed reservoirs or when a zero net volumetric flux is otherwise enforced, the opposing volume fluxes must be equal. Such flow configurations are encountered in volcanic conduits, groundwater wells, industrial pipelines, and other confined counter-current systems. However, even within the laminar regime, the density and viscosity ratios alone do not uniquely determine the cross-sectional flow configuration: concentric and eccentric core–annular structures, side-by-side stratified layers, and temporally unsteady switching between these states may all be realised. This multiplicity of admissible configurations is central to the present study, in which the internal interface may be either centred or displaced from the mid-plane of a vertical planar channel.

%\vspace{3mm}

Early studies of exchange in vertical tubes exposed a basic closure problem: zero net flux does not by itself determine the size or topology of the counter-current regions. \citet{HuppertHallworth2007} combined experiments, dimensional analysis and a steady axisymmetric Newtonian model, clarifying the balance among buoyancy, viscous stresses, and inertia. However, their core radius was selected through a maximum-flux hypothesis, and the model excluded asymmetric or fragmented states. \citet{BeckettEtAl2011} observed core–annular, side-by-side, and transitional regimes over a wide range of viscosity ratios. The viscosity contrast organised the regimes but did not uniquely select them, while neither the final topology nor the measured flux systematically maximised exchange. 
%\citet{Kerswell2011} considered steady Stokes solutions with circular-arc or circular interfaces and showed that maximum-flux states are generally asymmetric: side-by-side at lower viscosity ratios and eccentric core–annular at higher ratios. This demonstrated strong geometrical degeneracy, although the optimisation principle was heuristic and neither stability nor dynamical accessibility was established.
\citet{SuckaleEtAl2018} addressed non-uniqueness through a steady model and transient simulations, identifying thin- and thick-core branches and showing that boundary conditions and flow history can determine the realised state. Their results weaken universal selection rules based on maximum flux or minimum dissipation, although the analytical model does not independently determine the exchange flux. \citet{LongoEtAl2022} studied an axisymmetric non-Newtonian long drop rising in a vertical tube while a Newtonian liquid descended through the annulus. Under lubrication and zero-net flow assumptions, an energy balance yielded an asymptotic core radius that complied reasonably well with measured radii, velocities, and selected profiles. The approach provides a selection mechanism within a prescribed core–annular topology but not among competing topologies. \citet{CharabinFrigaard2024} examined the reversed rheological arrangement, with a viscoplastic fluid surrounding a lighter Newtonian finger. Yield stress could stop exchange or produce helical, disconnected-finger, and slug regimes, and the energy-based radius selection of \citet{LongoEtAl2022} did not produce an equilibrium in their parameter range. Collectively, these studies indicate that rheology, interface geometry, boundary conditions, and flow history all influence the observed state; no universal selection principle has emerged.

%\vspace{3mm}

The linear stability of vertical core–annular pipe flow has been investigated using increasingly comprehensive two-fluid models. \citet{Hickox1971} examined the fully developed concentric flow of two immiscible Newtonian fluids, accounting for density and viscosity contrasts, gravity, inertia, and surface tension. A long-wave asymptotic expansion demonstrated that stratification can destabilise the axisymmetric base flow; however, this analysis was restricted to long-wavelength perturbations and to configurations with a less viscous core fluid. \citet{ChenBaiJoseph1990} subsequently removed the long-wave limitation while retaining the effects of gravity, capillarity, inertia, and an imposed pressure gradient that could assist or oppose gravity. Their computations disentangled the respective roles of capillary forces, interfacial shear, and gravitational effects and identified a stability window corresponding to nearly ideal core–annular flow, which was later corroborated by free-fall experiments. In both studies, the base flow was assumed to be perfectly concentric and the analysis was primarily concerned with axisymmetric perturbations.

\citet{BaiChenJoseph1992} compared linear theory with vertical oil–water experiments in upward and downward flow. The theory reproduced wave celerities sufficiently well to classify regimes and relate them to bamboo and corkscrew waves, but it predicted wavelengths that were shorter than those observed and could not represent the three-dimensional corkscrew deformation. \citet{HuPatankar1995} therefore extended the theory to non-axisymmetric modes and showed that, for thin cores, a sinuous azimuthal disturbance can be more dangerous than the axisymmetric mode, providing a linear mechanism for corkscrew waves. These pipe-flow studies all employ a deformable interface with full kinematic and dynamic jump conditions; therefore, their instabilities are genuinely interfacial, unlike the prescribed-density approximation adopted here.

%\vspace{3mm}

The literature on vertical plane channels is more limited and especially pertinent. \citet{Renardy1987} analysed a steady, fully developed, pressure-driven three-layer Poiseuille flow between vertical parallel walls. A central Newtonian layer was symmetrically bounded by a second fluid, with deformable interfaces, density and viscosity contrasts, and surface tension. The stability problem contained snake and varicose interfacial modes, as well as modes related to single-fluid Poiseuille flow. Stability depended jointly on layer ordering, thickness, and properties: density and viscosity stratification could reinforce or oppose one another, and finite-wavelength modes could remain unstable even when the long-wave limit was stable. This is the closest sharp-interface predecessor to the present geometry, but it was restricted to a centred layer with symmetric outer regions and did not impose zero net flux.

\citet{Joo1995} considered two immiscible viscoelastic fluids in a vertical plane channel driven by gravity and pressure, with unequal wall temperatures. A long-wave evolution equation includes density, viscosity, and elasticity stratification, shear thinning, capillarity, and thermocapillarity, showing that Marangoni stresses may enhance or suppress interfacial instability. Its scope was limited by the lubrication approximation, the single-interface geometry, and the absence of a zero-net-flux constraint. \citet{SeginKondicTilley2006} examined a thin liquid film and a counter-current compressible gas in a narrow vertical or inclined channel. Under lubrication, ideal-gas, and isothermal assumptions, the gas compressibility was destabilising for very long waves but potentially stabilising at finite wavenumber. \citet{SchmidtEtAl2016} used a complete Orr--Sommerfeld formulation for a falling liquid film sheared by an upward gas in a vertical channel. The interfacial, shear, and internal modes could coexist and interact; direct numerical simulations confirmed initial linear growth but also revealed secondary instability, irregular waves, and overturning.

%\vspace{3mm}

These sharp-interface studies demonstrate how capillarity, density, and viscosity stratification, bulk shear, and interface motion generate multiple interacting instability mechanisms. The present study deliberately isolates a different problem. A lighter internal layer and two denser outer layers occupy a vertical plane channel; the two transverse interface positions are prescribed independently, allowing centred and off-centred base states. The net volume flux is zero, and the motion is driven solely by buoyancy. Within a Boussinesq-like approximation, viscosity is replaced by a uniform reference value, while density differences are retained only in the body-force term. The model, therefore, uses a single velocity and pressure field with a piecewise-constant buoyancy distribution.

The equilibrium velocity is obtained analytically for arbitrary interface positions and scales linearly with the Archimedes number. Varying the two positions generates a two-parameter family of symmetric and asymmetric buoyant flows. To our knowledge, the linear stability of this complete family has not previously been determined within a single-field, zero-net-flux Boussinesq-like formulation. The perturbation problem is reduced to two-dimensional disturbances, written as an Orr--Sommerfeld eigenvalue problem and solved by the Chebyshev--Galerkin method.

%\vspace{3mm}

The paper is organised as follows. Section~\ref{sec:model} introduces the mathematical model and the main assumptions, while \S~\ref{sec:boussinesq} presents the Boussinesq-like approximation and
the corresponding dimensionless formulation. The equilibrium buoyant flow is derived in \S~\ref{sec:equilibrium}. Section~\ref{sec:stability} is devoted to the linear stability analysis, including the application of Squire's theorem and the formulation of the Orr--Sommerfeld eigenvalue problem. The numerical method is described in \S~\ref{sec:numerics}, whereas the stability results for symmetric and asymmetric configurations are discussed in \S~\ref{sec:results}. Finally, the main conclusions are summarised in \S~\ref{sec:conclusions}.

%
%%%%%%%%%%%%%%%%%%%%%%%%%%%%%%%%%%%%%%%%%%%%%%
\section{Mathematical Model}\label{sec:model}
%%%%%%%%%%%%%%%%%%%%%%%%%%%%%%%%%%%%%%%%%%%%%%
%
Let us consider the two-phase flow in a vertical plane channel. The channel walls are considered to be rigid boundaries subject to no-slip conditions. 
%
%%%%%%%%%%%%%%%%%%%%%%%%%%
\subsection{Assumptions}
\begin{figure}
\centering
\includegraphics[width=0.35\textwidth]{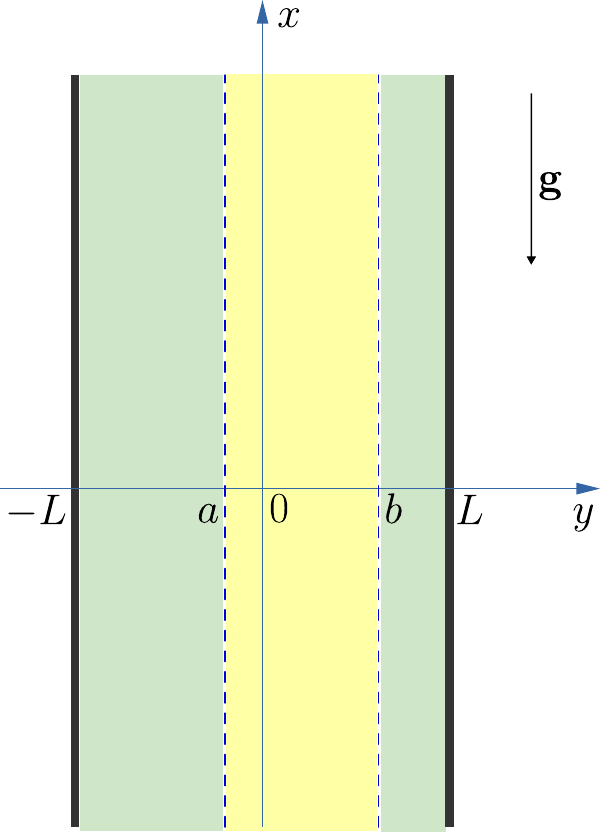}
\caption{\label{fig1}A sketch of the vertical channel with the yellow region denoting fluid 1 and the green regions denoting fluid 2.}
\end{figure}

%%%%%%%%%%%%%%%%%%%%%%%%%%
%
\begin{itemize}
\item The plane channel has a width $2L$, the vertical upward-orientated axis is $x$, and the horizontal axis in the span is $y$. The origin is taken so that $y=0$ is the midplane of the channel. Furthermore, the gravitational acceleration is $\vb{g} = \qty(-g, 0, 0)$.
\item There are two immiscible fluids labelled with $r=1, 2$. The fluid flowing in the internal core is identified with $r=1$, while that in the peripheral layer is denoted by $r=2$.
\item The flow regime is laminar, stationary, incompressible, and isothermal, with an overall flow rate equal to zero.
% \item The horizontal $z$ axis defines a direction of symmetry, so that both the velocity fields $\vb{u}_r$ and the pressure fields $p_r$ are independent of $z$ or equivalently, the flow is two-dimensional in the $(x,y)$ plane.
%\item The flow is fully-developed, so that the velocity fields have just a non-vanishing vertical component $\vb{u}_n = \qty(u_n, 0, 0)$ with $r=1,2$.
\end{itemize}
A sketch of the flow system can be found in figure~\ref{fig1}.

%
%%%%%%%%%%%%%%%%%%%%%%%%%%%%%%%%
\subsection{Governing equations}
%%%%%%%%%%%%%%%%%%%%%%%%%%%%%%%%
%
The local mass balance and momentum balance equations are given by
\eq{1}{
\div{\vb{u}_r} = 0, \nn
\rho_r \qty[ \pdv{\vb{u}_r}{t} + \qty(\vb{u}_r \vdot \grad) \vb{u}_r ] = - \grad{p_r} + \rho_r \vb{g} + \mu_r \laplacian{\vb{u}_r} \qc r = 1, 2,
}
where the densities $\rho_r$ and dynamic viscosities $\mu_r$ are constants, while the velocity fields and the pressure fields are denoted as  $\vb{u}_r$ and $p_r$, respectively.
Here, the equations of the inner core fluid $(r=1)$ are defined with $a < y < b$, where $a$ and $b$ can be positive or negative, provided that $|a|$ and $|b|$ are smaller than or equal to $L$. On the other hand, the outer layer equations  $(r=2)$ are defined with $-L < y < a$ and $b < y < L$. 

The boundary conditions at $y=\pm L$ are no-slip.
%
%\eq{2}{
%\vb{u}_2(x, \pm L, z, t) = 0.
%}
%
% By definition, equivalent flow cases are given by
%  %
% \eq{3}{
% \qty(a, b) \sim \qty(-b, -a).
% }
% %

%
%%%%%%%%%%%%%%%%%%%%%%%%%%%%%%%%%%%%%%%%
\section{A Boussinesq-like approximated scheme}\label{sec:boussinesq}
%%%%%%%%%%%%%%%%%%%%%%%%%%%%%%%%%%%%%%%%
%
The set of governing equations \eqref{1} can be approximated under the assumption that the quantities
\eq{4}{
\epsilon = \frac{\rho_2 - \rho_1}{\rho_0} \qc \delta = \frac{\mu_2 - \mu_1}{\mu_0},
}
are not too large. We further denote by $\rho_0$ and $\mu_0$ the average density and viscosity, respectively,
\eq{5}{
\rho_0 = \frac{1}{2L} \qty[ (b - a) \rho_1 + (2L - b + a) \rho_2] ,\nn
\mu_0 = \frac{1}{2L} \qty[ (b - a) \mu_1 + (2L - b + a) \mu_2].
}

Furthermore, we assume that the Archimedes number,
\eq{6}{
Ar = \frac{g L^3 \rho_0^2}{\mu_0^2}\, \epsilon ,
}
remains finite, while $|\epsilon| \ll 1$ and $|\delta| \ll 1$. Then, \eqref{1} can be approximated by introducing the overall velocity and pressure fields, $\qty(\vb{u}, p)$,
\eq{7}{
\div{\vb{u}} = 0, \nn
\rho_0 \qty[ \pdv{\vb{u}}{t} + \qty(\vb{u} \vdot \grad) \vb{u} ] = - \grad{p} + \rho(y) \vb{g} + \mu_0 \laplacian{\vb{u}} ,
}
with
\eq{8}{
\rho(y) = 
\begin{cases}
\rho_1 \qc a < y < b, \\
\rho_2 \qc -L < y < a \qand b < y < L . 
\end{cases}
}
%

%
%%%%%%%%%%%%%%%%%%%%%%%%%%%%%%%%%%%%%%%%%%%%%
\subsection{Approximated mathematical model}
%%%%%%%%%%%%%%%%%%%%%%%%%%%%%%%%%%%%%%%%%%%%%
%
Hereafter, we will focus on the case of a lighter fluid in the inner core so that $\rho_2 > \rho_1$, namely $\epsilon > 0$ and $Ar > 0$. We define
\eq{9}{
P = p + \rho_2 g x,
}
and
\eq{10}{
\hat\rho(y) = 
\begin{cases}
-\qty( \rho_2 - \rho_1 ) \qc a < y < b, \\
0 \qc -L < y < a \qand b < y < L . 
\end{cases}
}
Thus, \eqref{7} yields
\eq{11}{
\div{\vb{u}} = 0, \nn
\rho_0 \qty[ \pdv{\vb{u}}{t} + \qty(\vb{u} \vdot \grad) \vb{u} ] = - \grad{P} + \hat\rho(y) \vb{g} + \mu_0 \laplacian{\vb{u}} ,
}
The no-slip boundary conditions at $y=\pm L$ are given by
\eq{12}{
\vb{u}(x, \pm L, z, t) = 0.
}
%

%
%%%%%%%%%%%%%%%%%%%%%%%%%%%%%%%%%%%%
\subsection{Dimensionless equations}
%%%%%%%%%%%%%%%%%%%%%%%%%%%%%%%%%%%%
%
Let us introduce dimensionless scaling,
\eq{13}{
\frac{(x,y,z)}{L} \to (x,y,z) \qc \frac{t}{L/U_0} \to t \qc \frac{\vb{u}}{U_0} \to \vb{u} \qc \frac{P}{P_0} \to P  \qc \frac{\hat\rho(y) g L^2}{\mu_0 U_0} \to \hat\rho(y) ,
}
with
\eq{14}{
U_0 = \frac{\mu_0}{\rho_0 L} \qc P_0 = \frac{\mu_0^2}{\rho_0 L^2} .
}
Then, \eqref{11} and \eqref{12} can be rewritten as
\eq{15}{
\div{\vb{u}} = 0, \nn
\pdv{\vb{u}}{t} + \qty(\vb{u} \vdot \grad) \vb{u} = - \grad{P} - \hat\rho(y) \vu{e}_x + \laplacian{\vb{u}} ,\nn
\vb{u}(x, \pm 1, z, t) = 0.
}
The dimensionless function $\hat\rho(y)$ depends on the Archimedes number as
\eq{16}{
\hat\rho(y) = 
\begin{cases}
- Ar \qc A < y < B, \\
0 \qc -1 < y < A \qand B < y < 1 , 
\end{cases}
}
where $A = a/L$ and $B = b/L$.

\begin{figure}[ht!]
\centering
\includegraphics[width=0.7\textwidth]{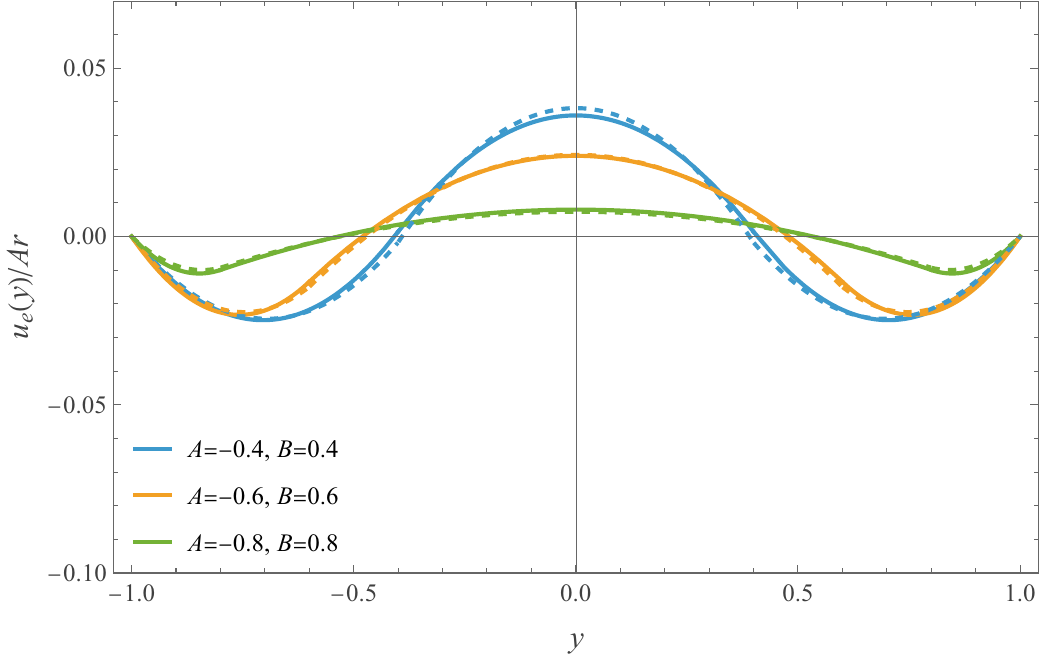}
\caption{\label{fig2}Plots of $u_e(y)/Ar$ versus $y$ for the symmetric cases, $A = - B$. The dashed lines represent the exact equilibrium solution with $\delta = 0.25$, described in Appendix~\ref{appendix1}.}
\end{figure}

\begin{figure}[ht!]
\centering
\includegraphics[width=0.7\textwidth]{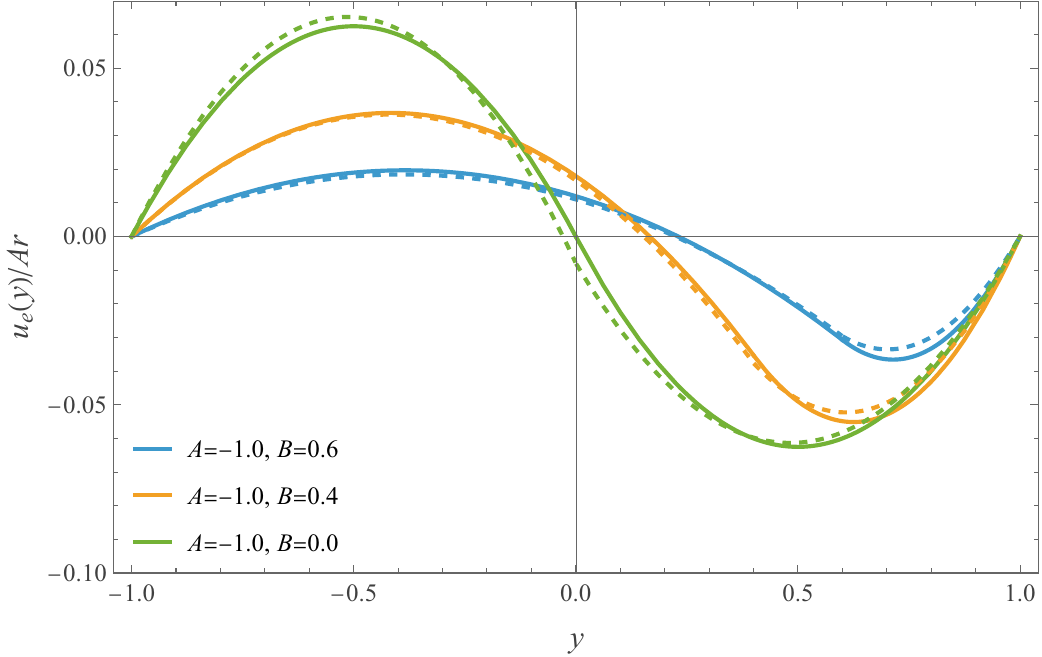}(a)\\[5mm]
\includegraphics[width=0.7\textwidth]{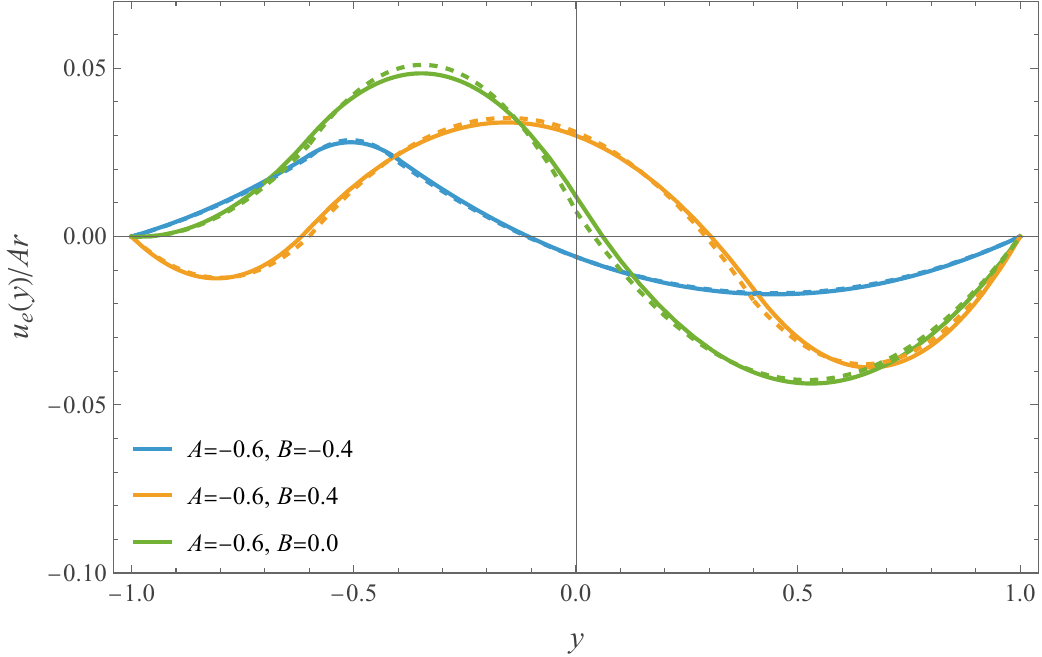}(b)
\caption{\label{fig3}Plots of $u_e(y)/Ar$ versus $y$ for asymmetric cases: two-layer asymmetric cases (a) and three-layer asymmetric cases (b). The dashed lines represent the exact equilibrium solution with $\delta = 0.25$, described in Appendix~\ref{appendix1}.}
\end{figure}

%
%%%%%%%%%%%%%%%%%%%%%%%%%%%%%%%%%%%%%%%%%%%%%%%%%%%%%%%%%%%%%%
\section{The equilibrium buoyant flow}\label{sec:equilibrium}
%%%%%%%%%%%%%%%%%%%%%%%%%%%%%%%%%%%%%%%%%%%%%%%%%%%%%%%%%%%%%%
%
There exists a stationary flow solution of \eqref{15} such that $\vb{u}_e = \qty(u_e(y) , 0 , 0)$ and $\grad{P_e} = \qty(K, 0, 0)$, where $K$ is a constant. Such a solution satisfies the ordinary differential equation
\eq{17}{
\dv[2]{u_e}{y} = K + \hat\rho(y), \nn
u_e(\pm 1) = 0.
}
The constant $K$ is determined by the constraint of a zero volume flow rate,
\eq{18}{
\int_{-1}^1 u_e(y)\, \dd y = 0.
}
Thus, the basic flow depends only on the three dimensionless parameters $A$, $B$, and $Ar$.

The solution of \eqref{17} satisfying \eqref{18} is given by
\eq{19}{
u_e(y) = \frac{Ar}{8} \Big\{(y+1) (A-B) \big[A^2 (y-1)+ (A + B) (B y - B +2)
- 3 y - 1\big]
\nn
\qquad +\; 4\,\theta(y-A) \left[(B-y)^2\, \theta(y-B)-(A-y)^2\right]\Big\} ,
}
with
\eq{20}{
K = \frac{Ar}{4} \qty(A-B)\qty(A^2+A B+B^2-3) .
}
In \eqref{19}, $\theta$ denotes Heaviside's unit step function,
\eq{21}{
\theta(s) = 
\begin{cases}
0 \quad\qfor s < 0,\\
1 \quad\qfor s \ge 0. 
\end{cases}
}

Moreover, by employing the unit step function, one can reformulate the expression of $\hat{\rho}(y)$ given by \eqref{16} as $\hat{\rho}(y) = - Ar\,\theta(y - A)\, \theta(B - y)$.

An obvious feature of \eqref{19} is that, in the limiting case where $A \to -1$ and $B \to 1$, then $u_e(y) \to 0$. In fact, since flow is caused only by buoyancy, there is no flow when the whole channel is occupied by a single fluid. For the same reason, $u_e(y) \to 0$ when $B \to A$.

The equilibrium buoyant flow $u_e(y)$ displays the symmetry
\eq{22}{
y \to -y \qc A \to -B \qc B \to -A \quad \Longrightarrow \quad u_e \to u_e  .
}
Due to this symmetry, we have the parametric equivalence $\qty(A, B) \sim \qty(A^*, B^*)$, with $A^*=-B$ and $B^*=-A$. 
%That is the dimensionless version of the equivalence \eqref{3}. 
Such an equivalence is important as it yields a simplification in the parametric space to be considered for the stability analysis of the equilibrium flow.

For completeness, Appendix~\ref{appendix1} provides a detailed description of the dimensionless formulation and the exact equilibrium solution obtained by relaxing the Boussinesq-like approximation employed so far.

Figures~\ref{fig2} and \ref{fig3} illustrate the velocity profiles $u_e(y)$ for some sample assignments of $A$ and $B$. Both symmetric cases (figure~\ref{fig2}) and asymmetric cases (figure~\ref{fig3}) are considered. The dependence on $Ar$ is encompassed by plotting the ratio $u_e(y)/Ar$. Figure~\ref{fig3}(a) is relative to flow regimes where $A=-1$, which means a two-layer configuration such that the special case $B=0$ defines the completely antisymmetric flow where $u_e(y)$ is an odd function. Figure~\ref{fig3}(b) shows asymmetric flow configurations with a three-layer configuration defined by $A=-0.6$.

In Figs.~\ref{fig2} and \ref{fig3}, the plots relative to the exact solution discussed in Appendix~\ref{appendix1} are presented for comparison as dashed curves. In particular, a fairly high value of $\delta$ is considered, namely $\delta=0.25$. This is a case where $\delta > 0$, or in other words, the core fluid 1 has a dynamic viscosity smaller than that of fluid 2. As the discrepancies relative to the Boussinesq-approximated velocity profiles turn out to be moderately small, one must consider that with $|\delta| < 0.25$, such discrepancies are even smaller and tend to zero when $\delta \to 0$, as expected.

We point out that the exact velocity profiles $u_e(y)$ satisfying the interface conditions in $y=A$ and $y=B$ must be continuous with a discontinuous first derivative at the interfaces~\citep{LongoEtAl2022}. That is a consequence of the difference in viscosity between the fluids, as discussed also in the Appendix~\ref{appendix1}. On the other hand, the approximate  velocity profiles $u_e(y)$ displayed in Figs.~\ref{fig2} and \ref{fig3} are continuous with a continuous first derivative, while the second derivative displays discontinuities at $y=A$ and $y=B$, as a consequence of \eqref{16} and \eqref{17}. 

The special case where $A = -1$ corresponds to a two-layer flow, with fluid 1 confined at $-1 < y < B$ and fluid 2 confined at $B < y < 1$. As the flow is buoyancy-induced, one expects that exchanging the layers occupied by the two fluids can be counterbalanced by reversing the direction of the $x$ and $y$ axes. %simultaneously reversing the direction of the $y$ axis and the local direction of the velocity.
Mathematically, this property for $A = -1$ is expressed by the symmetry,
\eq{22bis}{
y \to -y \qc B \to -B\quad \Longrightarrow\quad u_e \to - u_e  ,
}
implied by \eqref{19}.
This means, for instance, that
\eq{22ter}{
u_e\qty(y; A = -1, B = -0.2) = - u_e\qty(- y; A = -1, B = 0.2)  .
}
% 
%
%%%%%%%%%%%%%%%%%%%%%%%%%%%%%%%%%%%%%%%%%%%%%%%%%%%%%%%%%
\section{Linear stability analysis}\label{sec:stability}
%%%%%%%%%%%%%%%%%%%%%%%%%%%%%%%%%%%%%%%%%%%%%%%%%%%%%%%%%
%
We perturb the equilibrium buoyant flow by means of small-amplitude perturbations,
\eq{23}{
\vb{u} = \vb{u}_e + \tilde{\vb{u}} \qc \grad{P} = \grad{P}_e + \grad{\tilde{P}}.
}
Thus, using \eqref{15}, the dynamics of the perturbation fields, $\tilde{\vb{u}} = \qty(\tilde{u}, \tilde{v}, \tilde{w})$ and $\tilde{P}$, is governed by the linear equations,
\eq{24}{
\div{\tilde{\vb{u}}} = 0, \nn
\pdv{\tilde{u}}{t} + Ar\, U(y)\, \pdv{\tilde{u}}{x} + Ar\, \tilde{v}\, \DD U(y) = - \pdv{\tilde{P}}{x} + \laplacian{\tilde{u}} ,\nn
\pdv{\tilde{v}}{t} + Ar\, U(y)\, \pdv{\tilde{v}}{x} = - \pdv{\tilde{P}}{y} + \laplacian{\tilde{v}} ,\nn
\pdv{\tilde{w}}{t} + Ar\, U(y)\, \pdv{\tilde{w}}{x} = - \pdv{\tilde{P}}{z} + \laplacian{\tilde{w}} ,\nn
\tilde{\vb{u}}(x, \pm 1, z, t) = 0,
}
where the shortcut notation, 
\eq{25}{
\DD = \dv{}{y} \qc U(y) = \frac{u_e(y)}{Ar},
}
is adopted. We emphasise that $U(y)$ is independent of the Archimedes number.

%
%%%%%%%%%%%%%%%%%%%%%%%%%%%%%
\subsection{Squire's theorem}
%%%%%%%%%%%%%%%%%%%%%%%%%%%%%
%
Let us now express the perturbations as Fourier modes in the plane $(x,z)$,
\eq{26}{
\tilde{\vb{u}} = \vb{q}(y) \, e^{i\qty(k_x x + k_z z)} \, e^{\lambda t} \qc \tilde{P} = F(y) \, e^{i\qty(k_x x + k_z z)} \, e^{\lambda t} ,
}
where $\qty(k_x, 0, k_z)$ is the wave vector and $\lambda$ is the complex growth rate. Hence, \eqref{24} yields
\eq{27}{
\DD q_y + i\, k_x q_x + i\, k_z q_z = 0, \nn
\lambda\, q_x + i\, k_x Ar\, U(y)\, q_x + Ar\, q_y\, \DD U(y) = - i\, k_x F + \qty( \DD^2 - k^2 ) q_x ,\nn
\lambda\, q_y + i\, k_x Ar\, U(y)\, q_y = - \DD F + \qty( \DD^2 - k^2 ) q_y ,\nn
\lambda\, q_z + i\, k_x Ar\, U(y)\, q_z = - i\, k_z F + \qty( \DD^2 - k^2 ) q_z ,\nn
\vb{q}(\pm 1) = 0,
}
with $k$ the wavenumber, defined so that $k^2 = k_x^2 + k_z^2$. Employing the first equation \eqref{27}, the sum of the second \eqref{27} multiplied by $i\, k_x$, the third \eqref{27} derived with respect to $y$, and the fourth \eqref{27} multiplied by $i\, k_z$ yields
\eq{28}{
\qty( \DD^2 - k^2 ) F + 2\, i\, k_x Ar \, q_y\, \DD U(y) = 0 .
}
%
%Together with the third equation and the boundary conditions \eqref{27}, 
Equations \eqref{27} and \eqref{28} lead to
\eq{29}{
\qty( \DD^2 - k^2 ) F + 2\, i\, k_x Ar \, q_y\, \DD U(y) = 0 , \nn
\qty( \DD^2 - k^2 ) q_y - \lambda\, q_y - i\, k_x Ar\, U(y)\, q_y - \DD F = 0, \nn
q_y(\pm 1) = 0 \qc \DD q_y(\pm 1) = 0,
}
One can eliminate the unknown $F(y)$ from \eqref{29} and collapse the two ordinary differential equations of the second order into a single equation of the fourth order in the unknown $q_y(y)$, so that \eqref{29} can be rewritten as
\eq{30}{
\DD^4 q_y - \qty[2\,k^2 + \lambda + i\, k_x Ar\, U(y)] \DD^2 q_y 
\nn\qquad
+ \qty{k^2 \qty[k^2 + \lambda + i\, k_x Ar\, U(y)] + i\, k_x Ar\, \DD^2 U(y) } q_y = 0, \nn
q_y(\pm 1) = 0 \qc \DD q_y(\pm 1) = 0.
}
One can introduce the definition,
\eq{31}{
k\, Ar_s = k_x Ar .
}
It is evident that $Ar_s \le Ar$, where the equality holds only for the case of transverse modes, namely the modes with $k_z=0$. Employing $Ar_s$, the stability eigenvalue problem \eqref{30} becomes independent of the wave vector orientation in the $(x,z)$ plane,
\eq{32}{
\DD^4 q_y - \qty[2\,k^2 + \lambda + i\, k\, Ar_s\, U(y)] \DD^2 q_y 
\nn\qquad
+ \qty{k^2 \qty[k^2 + \lambda + i\, k\, Ar_s\, U(y)] + i\, k\, Ar_s\, \DD^2 U(y) } q_y = 0, \nn
q_y(\pm 1) = 0 \qc \DD q_y(\pm 1) = 0.
}
The solution of the eigenvalue problem \eqref{32} allows us to determine the neutral stability value of $Ar_s$ for a prescribed wavenumber $k$. In fact, neutral stability occurs when the real part of $\lambda$ is zero. For neutral stability, $Ar$ is at its minimum, {\em i.e.}, $Ar = Ar_s$, with transverse modes. Therefore, we can conclude that transverse modes, such that perturbations are $z$ independent, represent the most unstable modes. Actually, this statement is the thesis of Squire's theorem \citep{drazin2004hydrodynamic}.

%
%%%%%%%%%%%%%%%%%%%%%%%%%%%%%%%%%%%%%%%
\subsection{The Orr-Sommerfeld problem}
%%%%%%%%%%%%%%%%%%%%%%%%%%%%%%%%%%%%%%%
%
By employing Squire's theorem, we focus our linear stability analysis on transverse modes. Thus, \eqref{32} can be rewritten by setting
\eq{33}{
Ar_s = Ar \qc \lambda = - i\, c\, k \qc q_y(y) = \phi(y).
}
We can write
\eq{34}{
\DD^4 \phi - k \qty(2\,k - i\, c + i\, Ar\, U(y)) \DD^2 \phi 
\nn\qquad
+\, k \qty[k^2 \qty(k - i\, c + i\, Ar\, U(y)) + i\, Ar\, \DD^2 U(y) ] \phi = 0, \nn
\phi(\pm 1) = 0 \qc \DD \phi(\pm 1) = 0, \nn
\quad \qfor U(y) = \frac{1}{8} \Big\{(y+1) (A-B) \big[A^2 (y-1)+ (A + B) (B y - B +2)
- 3 y - 1\big]
\nn
\qquad\qquad +\; 4\,\theta(y-A) \left[(B-y)^2\, \theta(y-B)-(A-y)^2\right]\Big\} ,\nn
\quad \qand \DD^2 U(y) = \frac{1}{4} \qty(A-B)\qty(A^2+A B+B^2-3) - \theta(y - A)\, \theta(B - y) .
}
The real part of the complex constant $c\,k$ yields the angular frequency, $\omega$, of the normal mode, with $c$ representing the velocity of the complex phase. The imaginary part of $c$ is positive for linear instability, negative for linear stability, and zero for neutral stability. Furthermore, to derive the expression of $\DD^2 U(y)$ reported in \eqref{34}, we used \eqref{16}, \eqref{17}, \eqref{20}, \eqref{21}, and \eqref{25}.

%
%%%%%%%%%%%%%%%%%%%%%%%%%%%%%%%%%%%%%%%%%%%%%%
\section{Numerical method}\label{sec:numerics}
%%%%%%%%%%%%%%%%%%%%%%%%%%%%%%%%%%%%%%%%%%%%%%
%
For the solution of the Orr-Sommerfeld problem \eqref{34}, we will rely on the numerical method described in \citet{barletta2024alternative}. More precisely, we adopt a Galerkin method of order $N$. A shorthand notation for the fourth order differential equation \eqref{34} is
\eq{35}{
\mathcal{L}_1 \phi = c\, \mathcal{L}_2 \phi,
}
where $\mathcal{L}_1$ and $\mathcal{L}_2$ are linear differential operators given by
\eq{36}{
\mathcal{L}_1 = \DD^4 - k \qty[2\,k + i\, Ar\, U(y)] \DD^2 
+ k \qty{k^2 \qty[k + i\, Ar\, U(y)] + i\, Ar\, \DD^2 U(y) }, \nn
\mathcal{L}_2 = - i\, k \, \DD^2 + i\, k^3 .
}
The steps we take are the following:
\begin{itemize}
\item We expand the eigenfunction $\phi(y)$ in a series of basis functions $\psi_n(y)$ defined as
\eq{37}{
\psi_n(y) = (1 - y^2)^2\, T_n(y)\qc n = 0, 1,\ \ldots\ , N,
}
where $T_n(y)$ is the $n$th Chebyshev polynomial of the first kind \citep{abramowitz1965handbook}. Each basis function satisfies the boundary conditions \eqref{34} for $\phi(y)$.
\item The $N$th order approximation for the solution of \eqref{34} is given by
\eq{38}{
\phi(y) = \sum_{n=0}^{N} \alpha_n\, \psi_n(y),
}
where $\alpha_n$, with $n = 0, 1,\ \ldots\ , N$, are complex coefficients.
\item We substitute \eqref{38} into \eqref{35} and evaluate the scalar product with the basis function $\psi_m(y)$,
\eq{39}{
\sum_{n=0}^{N} \alpha_n \braket{\psi_m}{\mathcal{L}_1 \psi_n} = c\, \sum_{n=0}^{N} \alpha_n \braket{\psi_m}{\mathcal{L}_2 \psi_n},
}
where the scalar product $\braket{f_1}{f_2}$ is defined as
\eq{40}{
\braket{f_1}{f_2} = \int\limits_{-1}^{1} f_1(y)f_2(y)\, \dd y,
}
for any pair of integrable functions $f_1$ and $f_2$.
\end{itemize}
In fact, equation \eqref{39} can be rewritten as
\eq{41}{
\mathcal{A}\, \boldsymbol{\alpha} = c\, \mathcal{B}\, \boldsymbol{\alpha}
}
where $\mathcal{A}$ and $\mathcal{B}$ are $(N+1)\times(N+1)$ complex matrices and $\boldsymbol{\alpha} =\qty(\alpha_0, \alpha_1,\ \ldots\ , \alpha_N)$ is a $(N+1)$ dimensional complex vector. The generalised eigenvalue problem can be solved by standard algebraic techniques of any order $N$. This solution provides both the complex eigenvalue $c$ and the complex eigenvector $\boldsymbol{\alpha}$.

The numerical solution described above simplifies in the symmetric cases where $B = - A$. In fact, analysis can be limited to even basis functions, which means that the sum on the right-hand  side of \eqref{38} is restricted to even values of $n$. On the other hand, all non-negative integer values of $n$ have to be used for non-symmetric cases. We also mention that an alternative to Chebyshev polynomials is the use of Legendre polynomials  \citep{abramowitz1965handbook}, which means that \eqref{37} can be replaced by
\eq{42}{
\psi_n(y) = (1 - y^2)^2\, P_n(y)\qc n = 0, 1,\ \ldots\ , N,
}
where $P_n(y)$ is the $n$th Legendre polynomial.

The algorithm for determining the eigenvalue $c$ is implemented using \textsl{Octave} coding \citep{octave}. The integrations required for the matrix elements in \eqref{39} are performed by employing the Gauss-Legendre quadrature rule using the Golub-Welsch algorithm \citep{gander2014scientific}. In order to obtain satisfactory accuracy in numerical integrations, the number of quadrature points $N_q$ is kept proportional to the truncation order $N$, and in particular we set $N_q = 4 N$, which is definitely a conservative estimate for $N_q$. The algebraic solution of the eigenvalue problem \eqref{41} is obtained by employing the built-in function \texttt{eig} of \textsl{Octave}.

\section{Discussion of the results}\label{sec:results}
In the analysis of the onset conditions for linear instability, we will focus on different configurations defined by discrete values of $A$ and $B$ changing from $-1$ to $1$ in steps of $0.2$. Due to the general property $B > A$, the need for at least two different phase layers for a nontrivial equilibrium flow, and the symmetries expressed by \eqref{22} and \eqref{22bis}, the cases to be examined are much fewer than the 121 different pairs $(A,B)$ obtained from the discrete values of $A$ and $B$ mentioned above. Actually, as illustrated in table~\ref{tab0}, the independent cases to be investigated are only 25. One must be careful in applying such symmetries as that defined by \eqref{22}. $A \to -B$ and $B \to -A$ lead to an invariant eigenvalue, $c \to c$. On the other hand, the symmetry defined by \eqref{22bis}, $A=-1$, and $B \to -B$ leads to $c \to - \bar{c}$. Here, $\bar{c}$ stands for the complex conjugate of $c$. Then, if one applies \eqref{22bis}, the growth rate remains invariant, while the angular frequency changes sign. Physically, this means that the neutral stability curve remains invariant, but the vertically orientated propagation velocity of the neutrally stable modes reverses its direction with an invariant modulus. Exactly for this reason, if one applies the symmetry defined by \eqref{22bis} to the special case of the completely antisymmetric equilibrium flow, $A=-1$ and $B=0$, one is led to the equality $c = - \bar{c}$. In fact, in this special case, the transformation $A=-1$ and $B \to -B$ do not lead to a change. Thus, one can conclude that, for $A=-1$ and $B=0$, the real part of $c$ is zero, which means that the transition to instability is driven by non-travelling modes, namely modes with a zero angular frequency.

\begin{table*}[t]
  \centering
  \setlength{\tabcolsep}{0pt}
  \renewcommand{\arraystretch}{1.15}
  \begin{tabular}{c *{11}{C{1.1cm}}}
    \hline
    \diagbox{$A$}{$B$\quad} & $-1.0$ & $-0.8$ & $-0.6$ & $-0.4$ & $-0.2$ & $0.0$ & $0.2$ & $0.4$ & $0.6$ & $0.8$ & $1.0$ \\
    \hline
-1.0            &  \Gr &  \Bl &  \Bl &  \Bl &  \Bl & \Ve & \Ve & \Ve & \Ve & \Ve & \Gr \\ %\hline
-0.8            &  \Gr &  \Gr &  \Ve &  \Ve &  \Ve & \Ve & \Ve & \Ve & \Ve & \Ve & \Bl \\ %\hline
-0.6            &  \Gr &  \Gr &  \Gr &  \Ve &  \Ve & \Ve & \Ve & \Ve & \Ve & \Bl & \Bl \\ %\hline
-0.4            &  \Gr &  \Gr &  \Gr &  \Gr &  \Ve & \Ve & \Ve & \Ve & \Bl & \Bl & \Bl \\ %\hline
-0.2            &  \Gr &  \Gr &  \Gr &  \Gr &  \Gr & \Ve & \Ve & \Bl & \Bl & \Bl & \Bl \\ %\hline
0.0             &  \Gr &  \Gr &  \Gr &  \Gr &  \Gr & \Gr & \Bl & \Bl & \Bl & \Bl & \Bl \\ %\hline
0.2             &  \Gr &  \Gr &  \Gr &  \Gr &  \Gr & \Gr & \Gr & \Bl & \Bl & \Bl & \Bl \\ %\hline
0.4             &  \Gr &  \Gr &  \Gr &  \Gr &  \Gr & \Gr & \Gr & \Gr & \Bl & \Bl & \Bl \\ %\hline
0.6             &  \Gr &  \Gr &  \Gr &  \Gr &  \Gr & \Gr & \Gr & \Gr & \Gr & \Bl & \Bl \\ %\hline
0.8             &  \Gr &  \Gr &  \Gr &  \Gr &  \Gr & \Gr & \Gr & \Gr & \Gr & \Gr & \Bl \\ %\hline
1.0             &  \Gr &  \Gr &  \Gr &  \Gr &  \Gr & \Gr & \Gr & \Gr & \Gr & \Gr & \Gr \\
    \hline
  \end{tabular}
  \caption{Parameter grid for $A$ and $B$ (steps of 0.2): the grey cells yield physically precluded cases; the blue cells yield cases which can be targeted by employing the symmetries \eqref{22} and \eqref{22bis}; the green cells correspond to the independent cases to be examined.} \label{tab0}
\end{table*}

\begin{figure}[ht!]
\centering
\includegraphics[width=0.7\textwidth]{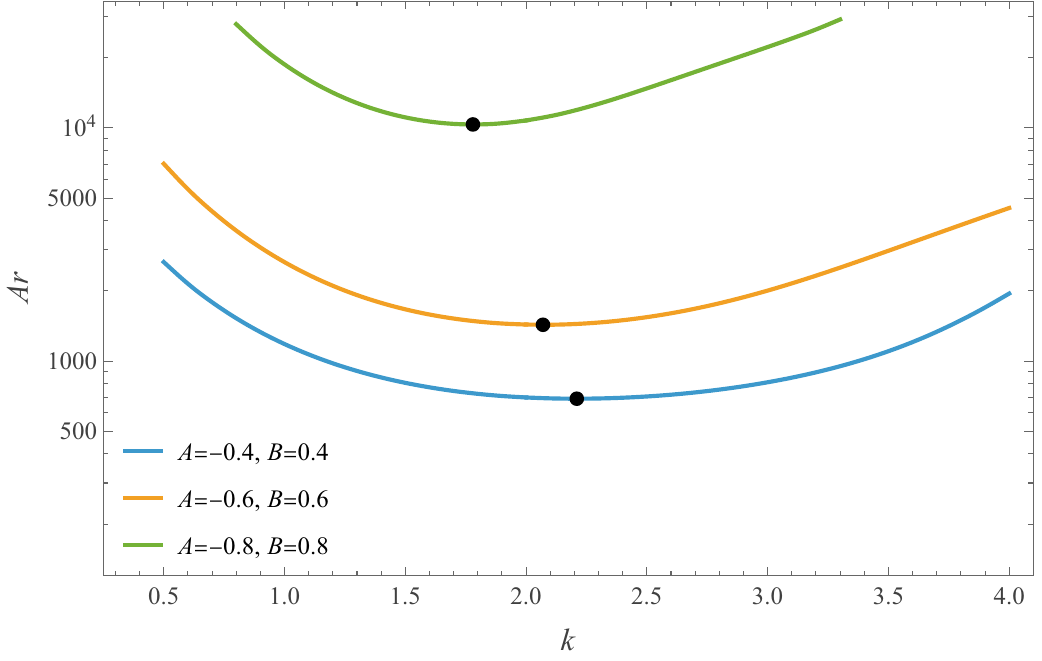}
\caption{\label{fig5}Neutral stability curves $Ar(k)$ for symmetric cases, $A = - B$. The black dots denote the minima of $Ar(k)$.}
\end{figure}

\begin{figure}[ht!]
\centering
\includegraphics[width=0.7\textwidth]{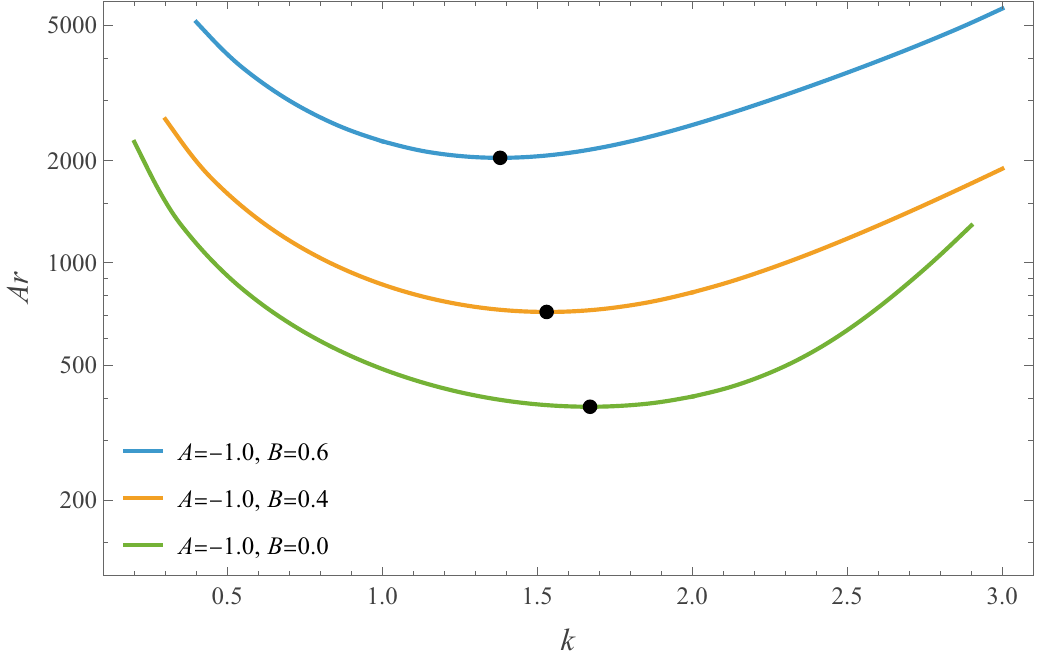}(a)\\[5mm]
\includegraphics[width=0.7\textwidth]{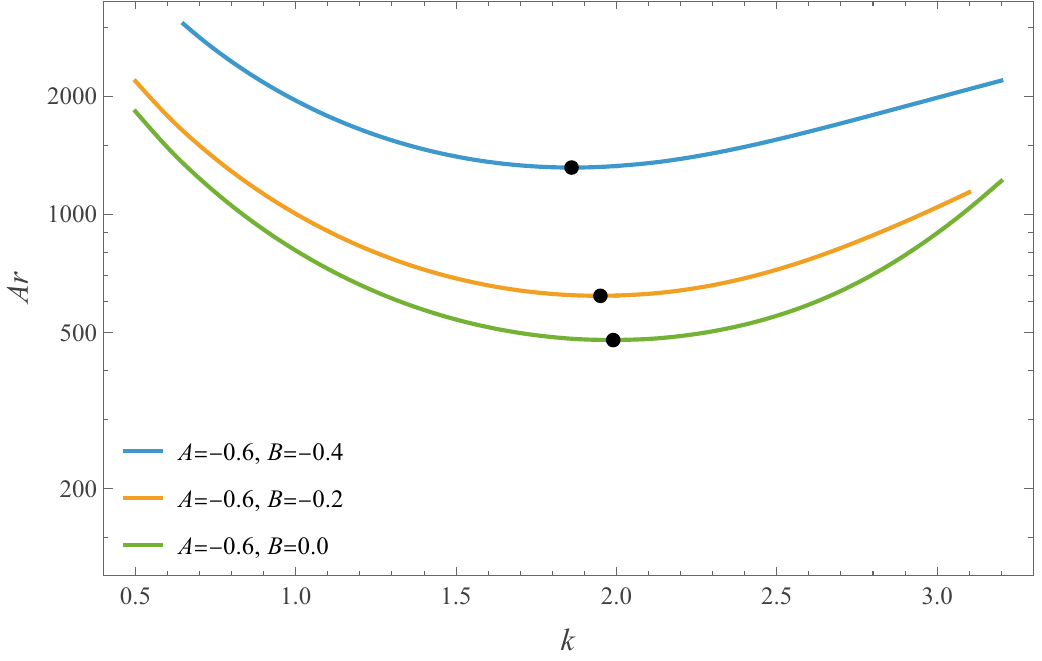}(b)
\caption{\label{fig6}Neutral stability curves $Ar(k)$: two-layer asymmetric cases (a) and three-layer asymmetric cases (b). The black dots denote the minima of $Ar(k)$.}
\end{figure}

\begin{table}[ht!]
\centering
\begin{tabular}{lllllllll}

& \multicolumn{1}{c}{$B=-0.6$}
& \multicolumn{1}{c}{$B=0$}
& \multicolumn{1}{c}{$B=0.6$} 
\\
\hline
$N=10$~~~ & $7.051\,121 -7.653\,730i$ ~~~& $-1.998\,785 +9.978\,089i$ ~~~& $-12.608\,177 -7.793\,327i$\\
$N=20$~~~ & $7.051\,132 -7.654\,188i$ ~~~& $-1.985\,449 +10.025\,111i$ ~~~& $-12.607\,902 -7.794\,023i$\\
$N=30$~~~ & $7.051\,136 -7.654\,194i$ ~~~& $-1.985\,403 +10.025\,415i$ ~~~& $-12.607\,915 -7.794\,029i$\\
$N=40$~~~ & $7.051\,137 -7.654\,195i$ ~~~& $-1.985\,401 +10.025\,430i$ ~~~& $-12.607\,916 -7.794\,030i$\\
$N=50$~~~ & $7.051\,137 -7.654\,195i$ ~~~& $-1.985\,401 +10.025\,432i$ ~~~& $-12.607\,916 -7.794\,030i$\\
$N=60$~~~ & $7.051\,137 -7.654\,195i$ ~~~& $-1.985\,401 +10.025\,432i$ ~~~& $-12.607\,916 -7.794\,030i$
 \\
\hline

\end{tabular}
\caption{Convergence test: dominant eigenvalues $c$ with different assignments of the truncation order $N$, for $A=-0.8$, $Ar=1000$, $k=1$ and different values of $B$.} \label{tab1}
\end{table}

%%%%%%%%%%%%%%%%%%%%%%%%%%%%%%%%%%%%%%

Table~\ref{tab1} illustrates a convergence test of the numerical method carried out for $A=-0.8$, $Ar=1000$, $k=1$, and $B = -0.6, 0, 0.6$. The convergence test is performed for an increasing truncation number $N$, where $(N+1)$ is the maximum number of terms in the Chebyshev polynomial approximation of the eigenfunction $\phi(y)$. The convergence test is carried out on the dominant eigenvalue $c$, which is the eigenvalue with the largest imaginary part. With $N$ increasing from $10$ to $60$, we see that the critical data converge within at least seven significant figures. Hence, in the following computations, we will use a truncation $N=40$ for all the cases examined. It is also important to mention that employing the Legendre polynomial basis \eqref{42}, instead of the Chebyshev polynomial basis, does not significantly affect the numerical results.%for a given $N$ within all the significant figures reported.

As mentioned in \S~\ref{sec:stability}, the neutral stability condition corresponds to a vanishing imaginary part of the dominant eigenvalue $c$. For the prescribed input values of $(A, B)$, the neutrally stable Archimedes number can be evaluated for every value of the wavenumber $k$. In this way, the function $Ar(k)$ that defines the neutral stability curve can be traced for every pair $(A, B)$. For a given wavenumber $k$, linear instability occurs whenever the Archimedes number is larger than the neutral stability threshold $Ar(k)$. Below such a threshold, we have linear stability.
Figures~\ref{fig5} and \ref{fig6} show such neutral stability curves for different pairs $(A, B)$. In particular, figure~\ref{fig5} is related to symmetric cases, where $A = -B$. On the other hand, figure~\ref{fig6} is related to asymmetric cases: figure~\ref{fig6}(a) shows cases with two layers $(A=-1)$, and figure~\ref{fig6}(b) illustrates cases with three layers $(A=-0.6)$. In each case considered in these figures, the shape of the neutral stability curves is upward concave, not very different from other buoyancy-induced instability cases such as the Rayleigh-B\'enard problem \citep[see, for instance,][]{drazin2004hydrodynamic}. At first glance, it seems that the neutral stability curves move downward in the $(k, Ar)$ plane as the width of the core phase, $B-A$, decreases. It looks fairly reasonable that a thinner core phase yields a potentially more unstable flow regime. However, this behaviour is detected in figure~\ref{fig5} and in figure~\ref{fig6}(a). On the other hand, figure~\ref{fig6}(b) suggests that this trend reverses when the width of the core phase becomes too small. A decreasing value of the width parameter $B-A$ may enhance the destabilising effect of the buoyancy force, while a too small width may lead to stabilisation of the flow for asymmetric configurations. 

The minimum of the neutral stability curve $Ar(k)$ is phenomenologically important, as it identifies the minimum $Ar$ that can lead to linear instability. As is well known, such a minimum value of $Ar$ is termed the critical Archimedes number and is denoted by $Ar_c$. Consequently, the corresponding wavenumber and angular frequency are denoted by $k_c$ and $\omega_c$, respectively.

%%%%%%%%%%%%%%%%%%%%%%%%%%%%%%%%%%%%%%
% \begin{table}
% \centering
% \def~{\hphantom{0}}
% \begin{tabular}{c|rrrrrrrrr}

% $\raisebox{-0.7mm}{$A$} \setminus \raisebox{0.1mm}{$B$}$
% & \multicolumn{1}{c}{$-0.8$}
% & \multicolumn{1}{c}{$-0.6$}
% & \multicolumn{1}{c}{$-0.4$}
% & \multicolumn{1}{c}{$-0.2$}
% & \multicolumn{1}{c}{$0$}
% & \multicolumn{1}{c}{$0.2$}
% & \multicolumn{1}{c}{$0.4$}
% & \multicolumn{1}{c}{$0.6$}
% & \multicolumn{1}{c}{$0.8$} \\[-2mm]

% \hline\\[-6mm]\\[-1mm]

% $-1$
% & \multicolumn{1}{c}{---}
% & \multicolumn{1}{c}{---}
% & \multicolumn{1}{c}{---}
% & \multicolumn{1}{c}{---}
% & $377$
% & $437$
% & $717$
% & $2040$
% & $19\,600$ \\

% $-0.8$
% & \multicolumn{1}{c}{---}
% & $2650$
% & $818$
% & $476$
% & $400$
% & $454$
% & $721$
% & $1880$
% & $10\,300$ \\

% $-0.6$
% & \multicolumn{1}{c}{---}
% & \multicolumn{1}{c}{---}
% & $1320$
% & $621$
% & $479$
% & $510$
% & $742$
% & $1430$
% & \multicolumn{1}{c}{---} \\

% $-0.4$
% & \multicolumn{1}{c}{---}
% & \multicolumn{1}{c}{---}
% & \multicolumn{1}{c}{---}
% & $1110$
% & $655$
% & $601$
% & $690$
% & \multicolumn{1}{c}{---}
% & \multicolumn{1}{c}{---} \\

% $-0.2$
% & \multicolumn{1}{c}{---}
% & \multicolumn{1}{c}{---}
% & \multicolumn{1}{c}{---}
% & \multicolumn{1}{c}{---}
% & $1240$
% & $750$
% & \multicolumn{1}{c}{---}
% & \multicolumn{1}{c}{---}
% & \multicolumn{1}{c}{---} \\

% \end{tabular}
% \caption{Critical values of the Archimedes number $Ar$ for given values of $A$ and $B$.} \label{tab2}
% \end{table}

\begin{table}
  \centering
  \setlength{\tabcolsep}{6pt}
  \begin{tabular}{l c c c c c c c c}
   % \toprule
    \diagbox{$A$}{$B$}\hspace{5mm} & $-0.6$ & $-0.4$ & $-0.2$ & $0.0$ & $0.2$ & $0.4$ & $0.6$ & $0.8$  \\
    \hline%\midrule
    $-1.0$      & $\cdots$ & $\cdots$ & $\cdots$ & $377.324$ & $437.281$ & $717.381$ & $2038.90$ & $19\,615.1$ \\
    $-0.8$      & $2648.76$ & $818.151$ & $476.019$ & $400.162$ & $453.948$ & $720.918$ & $1876.37$ & $10\,346.3$\\
    $-0.6$      & $\cdots$ & $1317.39$ & $620.563$ & $478.591$ & $510.129$ & $742.205$ & $1431.47$ & $\cdots$\\
    $-0.4$      & $\cdots$ & $\cdots$ & $1113.02$ & $654.621$ & $600.649$ & $689.711$ & $\cdots$ & $\cdots$ \\
    $-0.2$      & $\cdots$ & $\cdots$ & $\cdots$ & $1235.67$ & $749.862$ & $\cdots$ & $\cdots$ & $\cdots$ \\
   \hline %\bottomrule
  \end{tabular}
\caption{Critical Archimedes number, $Ar_c$, for given values of $A$ and $B$.} \label{tab2}
\end{table}
%%%%%%%%%%%%%%%%%%%%%%%%%%%%%%%%%%%%%%

%%%%%%%%%%%%%%%%%%%%%%%%%%%%%%%%%%%%%%
% \begin{table}
% \centering
% \begin{tabular}{c|ccccccccccc}

% $\raisebox{-0.7mm}{$A$} \setminus \raisebox{0.1mm}{$B$}$&   $-0.8$ & $-0.6$   & $-0.4$  & $-0.2$  & $0$  & $0.2$ & $0.4$ & $0.6$ & $0.8$  \\[-2mm]
%  \hline\\[-6mm]\\[-1mm]

% $-1$ &		---&	---&	---&	---&	$1.67$&	$1.63$&	$1.53$&	$1.38$&	$1.08$ \\

% $-0.8$ &		---&	$1.41$&	$1.60$&	$1.71$&	$1.75$&	$1.74$&	$1.68$&	$1.63$&	$1.78$  \\

% $-0.6$ & 	---&	---&	$1.86$&	$1.95$&	$1.99$&	$2.00$&	$2.02$&	$2.07$&	--- \\

% $-0.4$ & 	---&	---&	---&	$2.17$&	$2.21$&	$2.23$&	$2.21$&	---&	---  \\

% $-0.2$ & 		---&	---&	---&	---&	$2.35$&	$2.33$&	---&	---&	---  \\

% \end{tabular}\caption{Critical values of the wavenumber $k$ for given values of $A$ and $B$.} \label{tab3}
% \end{table}

\begin{table}
  \centering
   \setlength{\tabcolsep}{6pt}
  \begin{tabular}{l c c c c c c c c}
    %\toprule
    \diagbox{$A$}{$B$}\hspace{5mm} & $-0.6$ & $-0.4$ & $-0.2$ & $0.0$ & $0.2$ & $0.4$ & $0.6$ & $0.8$  \\
    \hline%\midrule
    $-1.0$      & $\cdots$ & $\cdots$ & $\cdots$ & $1.667\,6$ & $1.634\,0$ & $1.531\,9$ & $1.378\,7$ & $1.084\,4$ \\
    $-0.8$      & $1.414\,8$ & $1.599\,5$ & $1.710\,2$ & $1.750\,8$ & $1.735\,8$ & $1.678\,9$ & $1.629\,7$ & $1.781\,2$\\
    $-0.6$      & $\cdots$ & $1.862\,3$ & $1.948\,3$ & $1.985\,9$ & $2.003\,0$ & $2.023\,3$ & $2.069\,3$ & $\cdots$\\
    $-0.4$      & $\cdots$ & $\cdots$ & $2.167\,5$ & $2.207\,9$ & $2.228\,0$ & $2.208\,5$ & $\cdots$ & $\cdots$ \\
    $-0.2$      & $\cdots$ & $\cdots$ & $\cdots$ & $2.353\,8$ & $2.330\,9$ & $\cdots$ & $\cdots$ & $\cdots$ \\
    \hline%\bottomrule
  \end{tabular}
\caption{Critical wavenumber, $k_c$, for given values of $A$ and $B$.} \label{tab3}
\end{table}
%%%%%%%%%%%%%%%%%%%%%%%%%%%%%%%%%%%%%%

%%%%%%%%%%%%%%%%%%%%%%%%%%%%%%%%%%%%%%
% \begin{table}
% \centering
% \begin{tabular}{c|ccccccccccc}

% $\raisebox{-0.7mm}{$A$} \setminus \raisebox{0.1mm}{$B$}$&    $-0.8$ & $-0.6$   & $-0.4$  & $-0.2$  & $0$  & $0.2$ & $0.4$ & $0.6$ & $0.8$   \\[-2mm]
%  \hline\\[-6mm]\\[-1mm]
 
% $-1$ &		---&	---&	---&	---&	$0$&	$-4.35$&	$-11.3$&	$-30.0$&	$-126$  \\

% $-0.8$ &		---&	$24.1$&	$9.69$&	$3.56$&	$-0.512$&	$-4.72$&	$-11.4$&	$-27.7$&	$-72.1$  \\

% $-0.6$ & 		---&	---&	$5.06$&	$1.05$&	$-2.04$&	$-5.39$&	$-9.88$&	$-15.5$&	---  \\

% $-0.4$ & 		---&	---&	---&	$-1.35$&	$-3.21$&	$-4.89$&	$-6.55$&	---&	---  \\

% $-0.2$ & 		---&	---&	---&	---&	$-3.59$&	$-4.18$&	---&	---&	---  \\

% \end{tabular}\caption{Critical values of the angular frequency $\omega$ for given values of $A$ and $B$.} \label{tab4}
% \end{table}

\begin{table}
  \centering
   \setlength{\tabcolsep}{6pt}
  \begin{tabular}{l c c c c c c c c}
    %\toprule
    \diagbox{$A$}{$B$}\hspace{5mm} & $-0.6$ & $-0.4$ & $-0.2$ & $0.0$ & $0.2$ & $0.4$ & $0.6$ & $0.8$  \\
    \hline%\midrule
    $-1.0$      & $\cdots$ & $\cdots$ & $\cdots$ & $0$ & $-4.35$ & $-11.3$ & $-30.0$ & $-126$ \\
    $-0.8$      & $24.1$ & $9.69$ & $3.56$ & $-0.512$ & $-4.72$ & $-11.4$ & $-27.7$ & $-72.1$\\
    $-0.6$      & $\cdots$ & $5.06$ & $1.05$ & $-2.04$ & $-5.39$ & $-9.88$ & $-15.5$ & $\cdots$\\
    $-0.4$      & $\cdots$ & $\cdots$ & $-1.35$ & $-3.21$ & $-4.89$ & $-6.57$ & $\cdots$ & $\cdots$ \\
    $-0.2$      & $\cdots$ & $\cdots$ & $\cdots$ & $-3.59$ & $-4.17$ & $\cdots$ & $\cdots$ & $\cdots$ \\
    \hline%\bottomrule
  \end{tabular}
\caption{Critical angular frequency, $\omega_c$, for given values of $A$ and $B$.} \label{tab4}
\end{table}
%%%%%%%%%%%%%%%%%%%%%%%%%%%%%%%%%%%%%%

Tables~\ref{tab2}-\ref{tab4} report the critical values $Ar_c$, $k_c$, and $\omega_c$ for the physically independent pairs $(A, B)$ highlighted in green in table~\ref{tab0}. We recall that, in each case examined, $\omega_c$ is evaluated as the product of $k_c$ and the real part of the eigenvalue $c$ calculated when $k=k_c$ and $Ar=Ar_c$. We kept the precision of the numerical data reported in these tables to six significant digits for $Ar_c$, five significant digits for $k_c$, and only three significant digits for $\omega_c$. This difference in the number of digits reported is justified considering that $Ar_c$ is identified as an extremum point of the neutral stability function $Ar(k)$ (the minimum), and therefore its variability over a small interval of $k$ is very small. This is not the case for $\omega_c$, as $\omega(k)$ is subject to significant variation for a small change in $k$ close to $k_c$. 

The main physical information is gathered from table~\ref{tab2} where the data on $Ar_c$ are collected. We recover the effect of the width parameter for  fluid 1, $B-A$, on the stability/instability of the flow. The most stable case (the one with the largest $Ar_c$) turns out to be the one with the largest value of $B-A$ among the cases examined, namely $A=-1$ and $B=0.8$. This is physically reasonable, as this case is very close to $A=-1$ and $B=1$, where no flow occurs due to the absence of buoyancy, and hence no instability is possible, which means that $Ar_c \to \infty$. However, as already pointed out in the previous discussion on the neutral stability curves, the idea that $B-A$ determines how large $Ar_c$ is conflicts with the actual numerical data. This is evident from table~\ref{tab2} by checking, for instance, the cases where $(A,B)$ is equal to $(-1,0.6)$ and $(-0.8,0.8)$, which share the same value $B-A=1.6$, although the critical Archimedes numbers are utterly different. The unexpectedly complex role of $B-A$ arises from the influence of the concurrent width(s) of the fluid-2 layer(s) on the magnitude of the buoyancy-driven flow and, consequently, on the development of instability. Indeed, the smaller values of $Ar_c$ are in the central part of the table, and are relative to the cases where $B-A$ is not markedly small. The most unstable case turns out to be the completely antisymmetric flow $A=-1$ and $B=0$, with $Ar_c = 377.324$. As shown in table~\ref{tab4}, for the completely antisymmetric flow, the transition to instability is driven by non-travelling perturbation modes $(\omega_c=0)$. As already pointed out, this is definitely a consequence of the special invariance characterising this flow. In fact, the flow is invariant changing simultaneously $x \to -x$ and $y \to -y$. Stated differently, the instability of this flow is not influenced by the orientation of the perturbing wave relative to the $x$ axis, so that the critical mode is a non-travelling one.

%
%%%%%%%%%%%%%%%%%%%%%%%%%%%%%%%%%%%%%%%%%%%%%
\section{Conclusions}\label{sec:conclusions}
%%%%%%%%%%%%%%%%%%%%%%%%%%%%%%%%%%%%%%%%%%%%%
%
 Linear stability analysis of buoyancy-induced two-phase flow in a vertical plane channel has been carried out. The equilibrium flow has been assumed to be a parallel vertical flow, allowing for stratification where the lighter fluid occupies the central core, while the fluid having a larger density is stratified laterally, close to the boundaries. The special case where one of the lateral layers has a vanishing width, resulting in a two-layer flow, has also been included in the analysis. Unlike previous analyses available in the literature, this study is focussed on a purely buoyant flow that yields a vanishing net flow rate across the vertical channel. 

In general, the definition of the equilibrium buoyant flow and the linear stability analysis thereof depend on a plethora of different dimensionless parameters, such as the Archimedes number, the density ratio, and the viscosity ratio between the fluids, as well as the geometric parameters defining the widths and positions of the two or three layers of flowing phases. Such circumstances make a comprehensive analysis of the instability conditions quite a difficult task, as pointed out, for example, by \citet{Renardy1987}.

The approach pursued in this paper is based on a Boussinesq-like scheme that yields an approximate mathematical model of the equilibrium flow and the linear stability analysis. The approximation is meaningful in that the density difference between the phases divided by the average density is small enough, but the Archimedes number, $Ar$ proportional to the density difference, remains finite. This approximate scheme mimics the Boussinesq approximation usually employed to study the thermally-driven buoyant flows in single- or multiphase flow systems. In such non-isothermal flows, the leading parameter for the instability is the Rayleigh number, while in the present paper the leading parameter is the Archimedes number. Despite the  different physics underlying the types of buoyant flows explored, there is an intrinsic similarity between the Rayleigh number for natural convection and the Archimedes number for isothermal two-phase flows driven by buoyancy. 

According to the Boussinesq-like approximate scheme, both equilibrium flows and their instability are governed just by the Archimedes number $Ar$ and by a pair of dimensionless parameters, $A$ and $B$ with $B>A$, defining the positions where the interfaces between the two phases are located.

The following important results have been gathered through the linear stability analysis.
\begin{itemize}
    \item There are symmetries of the basic equilibrium flow that make different assignments of the pair $(A,B)$ equivalent for the transition to linear instability. One is the equivalence $(A,B) \sim (-B,-A)$. Another partial equivalence concerns the two-layer flow regimes where $A=-1$, predicting $B \sim -B$. The latter equivalence is termed partial as, with $A=-1$ and $B \to -B$, the equilibrium flow velocity changes sign.
    % The latter equivalence is only partial as, with $A=-1$ and $B \to -B$, the neutral stability curve is left invariant, but the angular frequency of the perturbation modes changes its sign.
    \item Squire's theorem holds for the linear stability analysis, meaning that such an analysis can be carried out by means of two-dimensional perturbations in the $xy$ plane, where $x$ is the vertical coordinate and $y$ is the horizontal transverse coordinate. 
    \item The special case of completely antisymmetric equilibrium flow, where vertical velocity is an odd function of transverse coordinate $y$, yields the most unstable conditions among the cases examined. In this special case, the transition to instability is triggered by non-travelling modes, that is, modes with a zero angular frequency and a non-zero wavenumber.
    \item The decreasing width of the core flow region, $B-A$, tends to destabilise the equilibrium flow for the symmetric cases where $B=-A$, except for small differences $B-A$. For asymmetric equilibrium flows, the position of the core layer with respect to the boundaries also tends to have a sensible influence on the instability conditions. 
\end{itemize}

The approximate method developed in this paper offers a novel viewpoint on the stability analysis of the equilibrium two-phase flow in a vertical plane channel. In fact, the Boussinesq-like scheme provides a simplified approach in which the buoyancy effect is pinpointed as the source of the instability. Thus, other important effects such as the interface discontinuity of the velocity gradient determined by the viscosity difference between the fluids or the surface tension come out as minor or higher order effects when compared to the buoyancy. This approximation makes sense in a special class of two-phase flows, much in the same way that the Boussinesq approximation adopted for thermal convection flows is reliable, as far as other effects, such as the compressibility of the fluid, can be neglected. This type of analysis can be further developed in the future, by focussing on two-phase vertical flow regimes more likely to be validated experimentally. An example widely explored in the literature is the core annular flow in a vertical pipe with circular cross-section.

\section*{Acknowledgements}
Three authors, A. Barletta, M. Celli, and V. Di Federico, acknowledge financial support from Alma Mater Studiorum Universit\`a di Bologna with grant RFO-2026.

\section*{Declaration of Interests}
The authors report no conflict of interest.

\bibliographystyle{model1-num-names}
\bibliography{biblio}

\appendix
\section{Deduction of the exact equilibrium flow solution}\label{appendix1}

If one does not employ the Boussinesq-like approximation defined in \S\ref{sec:boussinesq}, then one starts from \eqref{1} and defines
\eq{app1}{
P_r = p_r + \rho_2 g x \qc r = 1, 2.
}
Then, one can reformulate \eqref{1} in a dimensionless form as
\eq{app2}{
\div{\vb{u}_1} = 0, \nn
R_1\, \qty[ \pdv{\vb{u}_1}{t} + \qty(\vb{u}_1 \vdot \grad) \vb{u}_1 ] = - \grad{P_1} + Ar\, \vu{e}_x + M_1 \laplacian{\vb{u}_1} ,\qfor A < y < B, 
}
and
\eq{app3}{
\div{\vb{u}_2} = 0, \nn
R_2\, \qty[ \pdv{\vb{u}_2}{t} + \qty(\vb{u}_2 \vdot \grad) \vb{u}_2 ] = - \grad{P_2} + M_2 \laplacian{\vb{u}_2} ,\nn
\qfor -1 < y < A \qand B < y < 1,
}
where the dimensionless quantities $\vb{u}_{1,2}$ and $P_{1,2}$ are defined according to \eqref{13} and \eqref{14}. There are new dimensionless parameters, given by
\eq{app4}{
R_{1,2} = \frac{\rho_{1,2}}{\rho_0} \qc M_{1,2} = \frac{\mu_{1,2}}{\mu_0}.
}
Such parameters are linked to $\epsilon$ and $\delta$ defined in \eqref{4} so that, by employing also \eqref{5}, we can write
\eq{app5}{
\epsilon = R_2 - R_1 \qc \delta = M_2 - M_1 ,\nn
R_1 = 1 - \frac{1}{2}\ \qty(2 - B + A)\, \epsilon \qc R_2 = 1 + \frac{1}{2}\ \qty(B - A)
\,\epsilon , \nn
M_1 = 1 - \frac{1}{2}\ \qty(2 - B + A)\, \delta \qc M_2 = 1 + \frac{1}{2}\ \qty(B - A)
\,\delta.
}
Let us now consider the equilibrium flow solution where 
\eq{app6}{
\vb{u}_{1,2} = \qty(u_{e1,e2}(y) , 0 , 0) \qc \grad{P}_{1,2} = \qty(K, 0, 0).
}
Then, \eqref{app2} and \eqref{app3} yield
\eq{app7}{
M_1\ \dv[2]{u_{e1}}{y} = K - Ar \qc M_2\ \dv[2]{u_{e2}}{y} = K .
}
We note that the parameters $R_1$ and $R_2$ do not appear explicitly in the formulation as the inertial terms vanish for the equilibrium flow regime. The only relevant parameter embodying the density difference remains the Archimedes number. Equations \eqref{app7} are to be solved in the three-layer domain by using the six boundary and interface conditions,
\eq{app8}{
u_{e2}(-1) = 0 \qc u_{e2}(A) = u_{e1}(A) \qc M_2 \left.\dv{u_{e2}}{y}\right|_{y=A} = M_1 \left.\dv{u_{e1}}{y}\right|_{y=A}, \nn
u_{e1}(B) = u_{e2}(B) \qc M_1 \left.\dv{u_{e1}}{y}\right|_{y=B} = M_2 \left.\dv{u_{e2}}{y}\right|_{y=B} \qc u_{e2}(1) = 0 .
}
It is exactly the three-layer structure of the domain that justifies the need to determine six unknown integration constants defined by the general solution of \eqref{app7} and, hence, the need of the six conditions given by \eqref{app8}.

Once equations \eqref{app7} and \eqref{app8} are solved analytically, we get $u_{e1,e2}(y)$ expressed as second-degree polynomials in $y$. The final step is determining the dimensionless constant $K$ so that the average velocity through the channel cross-section is zero. This is achieved by imposing the constraint
\eq{app9}{
\int_{-1}^A u_{e2}(y)\, \dd y + \int_{A}^B u_{e1}(y)\, \dd y + \int_{B}^1 u_{e2}(y)\, \dd y = 0.
}
For the sake of brevity, we omit the final expressions of the dimensionless velocity in the three layers and that of $K$. We just emphasise that, exactly as for the Boussinesq approximated solution, both $u_{e1,e2}(y)/Ar$ and $K/Ar$ are independent of the Archimedes number. Actually, they parametrically depend only on $A$, $B$ and $\delta$. In the limit, $\delta \to 0$, one recovers the Boussinesq approximated solution given by \eqref{19} and \eqref{20}, for every pair $(A,B)$.

\end{document}